%% file: main.tex
\documentclass[conference]{IEEEtran}
\usepackage[utf8]{inputenc}

\usepackage[table,xcdraw]{xcolor}
\usepackage{booktabs} 
\usepackage{mathptmx}   
\usepackage{multirow}
\usepackage{amsfonts}
\usepackage{amsmath}
\usepackage{amsthm}
\usepackage{url}
\usepackage{xcolor}
\usepackage{graphicx}
\usepackage{cite}
\usepackage{booktabs}
\usepackage[ruled,linesnumbered]{algorithm2e}
\usepackage{mathtools} 
\usepackage{textcomp}
\usepackage{subcaption}
\usepackage{siunitx}
\usepackage{subfiles}
\usepackage{xspace}
\usepackage{bbm}
\usepackage{breqn}
\usepackage{graphicx}
\usepackage{blindtext}
\usepackage{enumerate}
\usepackage{amssymb}
\usepackage{pifont}
\usepackage{soul}

\usepackage{tikz}
\usepackage{tkz-tab}
\usetikzlibrary{automata,arrows,positioning,calc}
\usetikzlibrary{shapes,snakes}

\theoremstyle{definition}

\definecolor{OliveGreen}{cmyk}{0.64,0,0.95,0.40}

\definecolor{boristext}{rgb}{0.22, 0.44, 0.88}
\definecolor{boriscomments}{rgb}{0.88, 0.04, 0.04}

\def\BibTeX{{\rm B\kern-.05em{\sc i\kern-.025em b}\kern-.08em
		T\kern-.1667em\lower.7ex\hbox{E}\kern-.125emX}}
		
\IEEEoverridecommandlockouts

\begin{document}

\title{Stateless Reinforcement Learning for Multi-Agent Systems: the Case of Spectrum Allocation in Dynamic Channel Bonding WLANs}



\author{\IEEEauthorblockN{1\textsuperscript{st} Sergio Barrachina-Muñoz}
		\IEEEauthorblockA{\textit{CTTC}\\
			Barcelona, Spain \\
			sbarrachina@cttc.cat}
		\and
		\IEEEauthorblockN{2\textsuperscript{nd} Alessandro Chiumento}
		\IEEEauthorblockA{\textit{University of Twente}\\
			Enschede, Netherlands \\
			a.chiumento@utwente.nl}
		\and
		\IEEEauthorblockN{3\textsuperscript{th} Boris Bellalta}
		\IEEEauthorblockA{\textit{Universitat Pompeu Fabra}\\
			Barcelona, Spain \\
			boris.bellalta@upf.edu}
	}

\maketitle

\thispagestyle{plain}
\pagestyle{plain}

\begin{abstract}
    \subfile{sections/abstract}
\end{abstract} 

\subfile{sections/introduction}

\subfile{sections/ml}
\subfile{sections/mapping_rl}

\subfile{sections/rl_models}
\subfile{sections/use_case}

\subfile{sections/future}

\section*{Acknowledgments}

\small{The work of Sergio Barrachina-Muñoz and Boris-Bellalta was supported in part by Cisco, WINDMAL under Grant PGC2018-099959-B-I00 (MCIU/AEI/FEDER,UE) and Grant SGR-2017-1188. Alessandro Chiumento is partially funded by the InSecTT project (https://www.insectt.eu/) which has   received   funding from   the   ECSEL   Joint   Undertaking   (JU)   under   grant agreement  No  876038.}
\bibliographystyle{IEEEtran}
\bibliography{bib}


\end{document}

%% file: sections/abstract.tex
Spectrum allocation in the form of primary channel and bandwidth selection is a key factor for dynamic channel bonding (DCB) wireless local area networks (WLANs). To cope with varying environments, where networks change their configurations on their own, the wireless community is looking towards solutions aided by machine learning (ML), and especially reinforcement learning (RL) given its trial-and-error approach. However, strong assumptions are normally made to let complex RL models converge to near-optimal solutions. Our goal with this paper is two-fold: justify in a comprehensible way why RL should be the approach for wireless networks problems like decentralized spectrum allocation, and call into question whether the use of complex RL algorithms helps the quest of rapid learning in realistic scenarios. We derive that stateless RL in the form of lightweight multi-armed-bandits (MABs) is an efficient solution for rapid adaptation avoiding the definition of extensive or meaningless RL states.


%% file: sections/introduction.tex
\section{Introduction} \label{section:introduction}

State-of-the-art applications like virtual reality or 8K video streaming are urging next-generation wireless local area networks (WLANs) to support ever-increasing performance demands. To enhance spectrum usage in WLANs, channel bonding at the 5 GHz band was introduced in 802.11n-2009 for bonding up to 40 MHz, and further extended in 802.11ac/ax and in 802.11be to bond up to 160 and 320 MHz, respectively. 
In this regard, dynamic channel bonding (DCB) is the most flexible standard-compliant policy since it bonds all the allowed idle secondary channels to the primary channel at the backoff termination~\cite{barrachina2019dynamic}.

By operating in unlicensed bands, anybody can create a new WLAN occupying one (or multiple) channels. Further, transmissions are initiated at random as mandated by the distributed coordination function. These factors generate uncertain contention and interference that are exacerbated when implementing channel bonding~\cite{deek2013intelligent}. So, despite the 5 GHz Wi-Fi band remains generally under-utilized~\cite{barrachina2021wi}, there are peaks in the day where parts of the spectrum get crowded. This leads, especially in uncoordinated high-density deployments, to different well-known problems like hidden and exposed nodes, ultimately limiting Wi-Fi's performance.

To keep a high quality of service in such scenarios, WLANs must find satisfactory configurations \textit{as soon as possible}.
In this regard, many heuristics-based works on custom spectrum allocation solutions have been proposed (e.g., \cite{xu2012opportunistic, zheng2014stochastic}). Heuristics are low-complex problem-solving methods that work well in steady scenarios. However, their performance is severely undermined in dynamic scenarios since they rely on statistics only from recent observations that tend to be outdated when new configurations are applied.

To overcome heuristics' limitations, we first motivate that machine learning (ML) can tackle the joint problem of primary and secondary channels allocation in DCB WLANs. Then, we address why supervised and unsupervised learning are not suitable for the problem given the need for fast adaptation in uncertain environments where it is not feasible to train enough to generalize. As such, we envision reinforcement learning (RL) as the only ML candidate to quickly adapt to particular (most of the times unique) scenarios.
We then justify why stateless RL formulations and, in particular, multi-armed bandits (MABs) are better suited to the problem over temporal difference variations like Q-learning. As well, we call into question the usefulness of deep reinforcement learning (DRL) due to the unfeasible amount of data needed to learn \cite{de2019deep}.

Finally, we present a self-contained example to compare the performance of a MAB based on $\epsilon$-greedy against the well-known Q-learning (relying on states), and a contextual variation of the $\epsilon$-greedy MAB. Extensive simulations show that the regular MAB outperforms the rest both in learning speed and mid/long-term performance. These results suggest that deploying lightweight MABs in uncoordinated deployments like domestic access points (APs) would boost performance.

%% file: sections/ml.tex
\section{A change of paradigm towards reinforcement learning in WLANs}

A heuristic is any problem-solving method that employs a practical, flexible, shortcut that is not guaranteed to be optimal but is good enough for reaching a short-term goal. Heuristics are then used to find quickly satisfactory solutions in complex systems where finding an optimal solution is infeasible or impractical~\cite{pearl1984heuristics}.
%
Heuristics are tied to the validity of recent observations, and such depend on the speed at which environment changes. Moreover, they do not consider the performance of previous actions, so they do not learn. Then, in uncoordinated, high-density deployments, where multiple DCB WLANs serving numerous stations (STAs) adapt their spectrum configurations on their own, designing accurate hand-crafted spectrum allocation heuristics is unfeasible.
Learning from past experience represents then a very strong alternative. The capability of ML to go beyond rule of thumb strategies by automatically learning (and adapting) to (un)seen situations can cope with heterogeneous scenarios~\cite{zappone2019wireless}.
From the three main types of ML, we envision RL as the most suitable one for adapting to scenarios where no training data is available. Practically, in RL an agent tries to learn how to behave in an environment by performing actions and observing the collected rewards. At a given iteration $t$, action $a$ (e.g., switch primary channel) results in a reward observation (e.g., throughput) drawn from a reward distribution $r_t(a) \sim \theta_a$.
Even though observations can be misleading in RL as well, agents rely on the learning performed through historical state/action-reward pairs, generating action selections policies beyond current observations.

As for the alternative main ML types, supervised learning (SL) learns a function that maps an input $x$ to an output $y$ based on example input-output labeled pairs $(x_i, y_i)$. One way to apply SL in our problem would be to try to learn the \textit{true} WLAN behavior$f(x) = y$, mapping the WLAN scenario (e.g., nodes' locations, traffic loads, spectrum configurations, etc.) $x$ to the real performance (e.g., throughput of each WLAN)  $y$ through an estimate $h_\theta(x) = \hat{y}$, where $h_\theta(x)$ is the learned function. Unfortunately, generating an acceptable dataset is an arduous task that may take months. Besides, the input domain of $x$ is multi-dimensional (with even categorical dimensions like the primary channel), so function $f$ representing WLANs' \textit{behavior} is so complex that learning an accurate estimate $h_\theta$ for realistic deployments is unfeasible.

Finally, in unsupervised learning (UL), there are only inputs $x$ and no corresponding output variables. UL's goal is to model the underlying structure or distribution of the observed data. Similarly to SL, the underlying structure of the observed data in the problem at issue is so complex that any attempt to model it through UL will be most likely fruitless. Hence, we believe that following an RL approach is preferable: adapt from scratch no matter the environment's intrinsic nature.

%% file: sections/mapping_rl.tex
\section{Mapping the problem to RL}


\subsection{Attributes, actions, and states}    \label{sec:actions}

While it is clear that the primary channel is critical to the WLAN performance since it is where the backoff procedure runs, the maximum bandwidth is significant as well. Indeed, once the backoff expires, the channel bonding policy and maximum allowed bandwidth will determine the bonded channels. Hence, limiting the maximum bandwidth may be appropriate to WLANs in order to reduce adverse effects like unfavorable contention or hidden nodes.
Thus, we consider two configurable attributes each agent-empowered AP can modify during the learning process: the primary channel where the backoff procedure is executed, $p \in \{1,2,3,4\}$, and the maximum bandwidth (in number of basic channels), $b \in \{1, 2, 4\}$. With DCB, the transmitter can adapt to the sensed spectrum on a per-frame basis. So, the bandwidth limitation just sets an upper bound on the number of basic channels to bond.
Accordingly, each WLAN has an action space $\mathcal{A}$ of size $4 \times 3 = 12$, where actions (or spectrum configurations) have the form $a = (p,b) \in \mathcal{A}$. Should we consider a central single agent managing multiple WLANs, the central action space would increase exponentially with the number of WLANs.

Finally, a \emph{state} $s$ is a representation of the world that guides the agent to select the adequate action $a$ by fitting $s$ to the RL policy. Since most of the time the agent cannot know the whole system in real-time, it may rely on partial and delayed representations of the environments through some discretization function mapping observations to states. Besides, since observations are application and architecture-dependent -- given they may be limited to local sensing capabilities or, on the contrary, be shared via a central controller -- the states' definition is also strictly dependent on the RL framework.

\subsection{Problem definition}

Our goal for this use-case is to improve the network throughput. The key then is to let the WLANs find (learn) the best actions $(p,b)$, where \textit{best} depends on the problem formulation itself.
We define the throughput satisfaction $\sigma$ of a WLAN $w$ at iteration $t$ simply as the ratio of throughput $\Gamma_{w,t}$ to generated traffic $\ell_{w,t}$ in that iteration, i.e., $\sigma_{w,t} = \Gamma_{w,t}/\ell_{w,t}$.
So, $\sigma_{w,t} \in [0,1], \forall w,t$. A satisfaction value $\sigma=1$ indicates that all the traffic has been successfully received at the destination.

Once the main performance metric has been defined, we can formulate the reward function $R$, the steering operator of any RL algorithm. For simplicity, we define $R \triangleq \sigma$ given that the throughput satisfaction is bounded per definition between 0 and 1, which is convenient for guaranteeing convergence in many RL algorithms.
Regardless of the reward function $R$, we aim at maximizing the cumulative reward \textit{as soon as possible}, $G = \sum_{t=1}^{T}r_t$,
%
where $T$ is the number of iterations available to execute the learning, and $r_{t}$ is the reward at iteration $t$. Let us emphasize the complexity of the problem by noting that $r_{t}$ depends on the reward definition $R$, on the action $a_t$ selected at $t$, and also on the whole environment (a.k.a, world) at $t$.

%% file: sections/rl_models.tex
\section{Reinforcement learning models}


Different ML solutions for the spectrum allocation problem, especially in RL, have been proposed in recent years.
However, certain assumptions hinder accurate evaluation. For instance, most papers consider synchronous time slots (e.g., \cite{chen2020research}). Also, some papers define a binary reward, where actions are simply good or bad (e.g., \cite{zhong2018actor}). This, while easing complexity, is not convenient for continuous-valued performance metrics like throughput. Further, most of the papers consider fully backlogged regimes (e.g., \cite{nakashima2020deep}), thus overlooking the effects of factual traffic loads. Lastly, many papers in the literature provide RL solutions without a clear justification of how the states, actions, and rewards are designed, and turn out to be only applicable to the problem at hand.




The most common formulations of RL rely on states, which are mapped through a policy $\pi$ to an action $a$, i.e., $\pi(s) = a$. However, 
a key benefit of stateless RL is precisely the absence of states, which eases its design by relying only on action-reward pairs. The most representative stateless RL model for non-episodic problems is the Multi Armed Bandit (MAB).
In MABs an action must be selected between fixed competing choices to maximize expected gain. Each choice's properties are only partially known at the time of selection and may become better understood as time passes~
\cite{sutton2018reinforcement}. MABs are a classic RL solution that exemplifies the exploration-exploitation trade-off dilemma.


The classical RL formulation is through temporal difference (TD) learning, a combination of Monte Carlo and dynamic programming. Like Monte Carlo, TD samples directly from the environment; like dynamic programming, 
TD perform updates based on current estimates (i.e., they \textit{bootstrap})\cite{sutton2018reinforcement}.
There are two main TD methods: state–action–reward–state–action (SARSA) and Q-learning. SARSA and Q-learning work similarly, assigning values to state-action pairs in a tabular way. The key difference is that SARSA is on-policy, whereas Q-learning is off-policy and uses the maximum value over all possible actions,
\begin{equation}    \label{eq:qlearning}
    \begin{split}
        Q(s_t,a_t) & := Q(s_t,a_t)\\
        & + \alpha_t \big(r_t + \gamma_t \max_{a'} Q(s_{t+1,a'}) - Q(s_t,a_t)\big),
    \end{split}
\end{equation}
where $Q$ is the Q-value, $\alpha$ is the learning rate , and $\gamma$ is the discount factor.


Finally RL's combination with deep learning results in deep reinforcement learning (DRL), where the policy or algorithm uses a deep neural network rather than tables.
The most common form of DRL is a deep learning extension of Q-learning, called deep Q-network (DQN). So, as in Q-learning, the algorithms still work with state-action pairs, conversely to the MAB formulation. While DQN is much more complex than Q-learning, it compensates for the very slow convergence of tabular Q-learning when the number of states and actions increases. That is, it is so costly to keep gigantic tables that it is preferable to use DNNs to approximate such a table.


For the problem of spectrum allocation in uncoordinated DCB WLANs, we anticipate MABs as the best RL formulation. The reason lies in the fact that meaningful states are intricate to define and their effectiveness heavily depends on the application and the type of scenario under consideration. In the end, a state is a piece of information that should help the agent to find the optimal policy. \emph{If the state is meaningless, or worst, misleading, it is preferable not to rely on states and go for a stateless approach}. In summary, when there is not plenty of time to train TD methods like Q-learning, it seems much better to go for a pragmatic approach: do gamble and find a satisfactory configuration as soon as possible, even though you will be most likely renouncing to an optimal solution.

%% file: sections/use_case.tex
\section{A use case on spectrum management}


\subsection{System model and deployment}

\begin{figure}[t]
	\centering
	\begin{subfigure}{0.255\textwidth}
		\includegraphics[width=\textwidth]{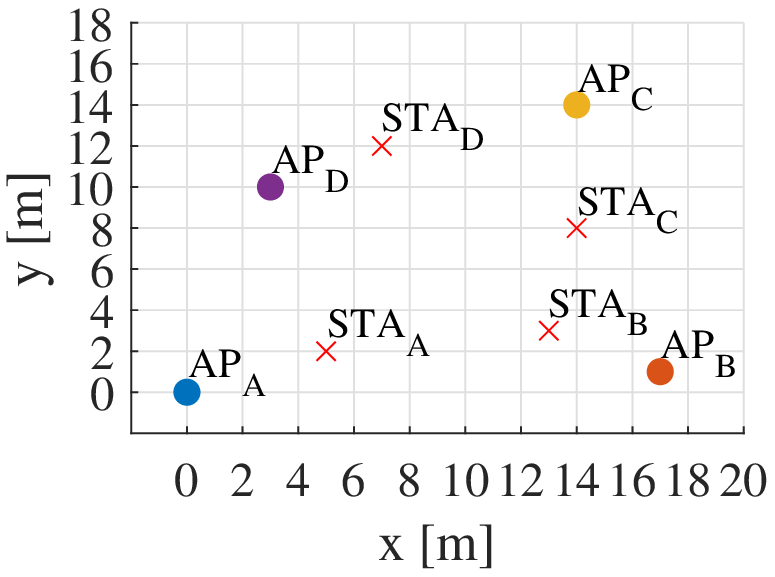}
		\caption{Location of nodes.}
		\label{fig:main_deployment}
	\end{subfigure}
	~
	\begin{subfigure}{0.19\textwidth}
		\includegraphics[width=\textwidth]{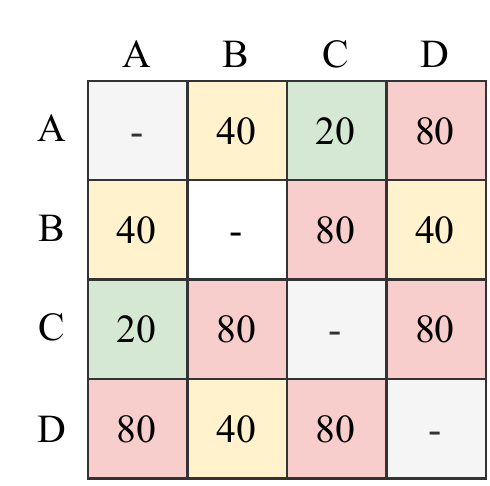}
		\caption{Interference matrix.}
		\label{fig:interference_matrix}
	\end{subfigure}
	
	\caption{Toy deployment of the self-contained dataset.}
	\label{fig:main_deployment_complete}
\end{figure}

We study the deployment illustrated in Figure~\ref{fig:main_deployment}, consisting of 4 potentially overlapping agent-empowered BSSs (with one AP and one STA each) in a system of 4 basic channels. The TMB path-loss model is assumed~\cite{adame2019tmb}, and spatially distributed scenarios are considered. Namely, we cover scenarios where the BSSs may not all be inside the carrier sense range of each other. Therefore, the typical phenomena of home Wi-Fi networks like flow in the middle and hidden/exposed nodes are captured~\cite{barrachina2019dynamic}.
The primary channel can take any of the 4 channels in the system, $p\in \{1,2,3,4\}$, and the maximum bandwidth fulfills the 802.11ac/ax channelization restrictions, $b\in\{1,2,4\}$, for 20, 40, and 80 MHz bandwidths. So, the total number of global configurations when considering all BSSs to have constant traffic load raises to \mbox{($4 \times 3)^4 = 20,736$}. Notice that even a petite deployment like this leads to a vast number of possible global configurations.

The interference matrix Figure~\ref{fig:interference_matrix} indicates the maximum bandwidth in MHz that causes two APs to overlap given the power reduction per Hertz when bonding. So, we note that this deployment is complex in the sense that multiple different one-to-one overlaps appear depending on the distance and the transmission bandwidth in use, leading to exposed and hidden node situations hard to prevent beforehand. For instance, $\text{AP}_{\text{A}}$ and $\text{AP}_{\text{C}}$ only overlap when using 20 MHz, whereas $\text{AP}_{\text{A}}$ and $\text{AP}_{\text{D}}$ always overlap because of their proximity.

We assume all BSSs perform DCB, being the maximum number of channels to be bonded limited by the maximum bandwidth attribute $b$.
Simulations of each global configuration are done through the Komondor wireless networks simulator~\cite{barrachina2019komondor}. The adaptive RTS/CTS mechanism introduced in the IEEE 802.11ac standard for dynamic bandwidth is considered and data packets generated by the BSSs follow a Poisson process with a mean duration between packets given by $1/\ell$, where $\ell$ is the mean load in packets per second. In particular, we consider data packets of 12000 bits and all BSSs having a high traffic load $\ell = 50$ Mbps.
Once the dataset is generated, we simulate the action selection of the agents to benchmark the performance of different RL models.\footnote{The dataset and Jupyter Notebooks for simulating the multi-agent behavior in uncoordinated BSSs is available at \url{https://github.com/sergiobarra/MARLforChannelBondingWLANs}.} 
\subsection{Benchmarking RL approaches}

The main drive behind the adoption of MABs is to avoid states entirely. This way, there is no need to seek meaningful state definitions, and the learning can be executed more quickly since only actions and rewards are used to execute the policy. Nonetheless, in this experiment, we provide a quantitative measure on Q-learning performance (relying on states) and contextual MABs (relying on contexts).

\subsubsection{State spaces}

\begin{figure}[h]
    \centering
    \includegraphics[width=0.35\textwidth]{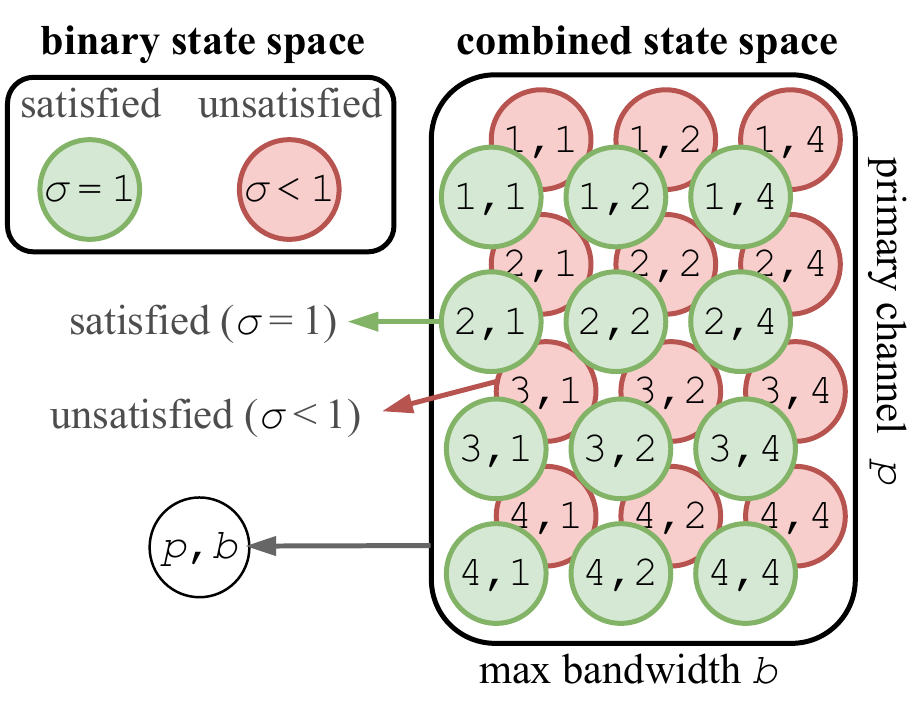}
    \caption{Binary and combined state/context spaces.}
    \label{fig:state_spaces}
\end{figure}

Apart from the default stateless approach of regular MABs, we propose two state (or context) space definitions as shown in Figure~\ref{fig:state_spaces}. The smaller state space is binary and represents the throughput satisfaction, i.e., it is composed of the satisfied and unsatisfied states. That is, if all the packets are successfully transmitted, the BSS is satisfied ($\sigma=1$), and unsatisfied otherwise ($\sigma<1$).
The larger state space is 3-dimensional, and combines the binary throughput satisfaction and the action attributes (primary $p$ and max bandwidth $b$). So, the size raises from 2 to 24 states, corresponding to the combination of 2 binary throughput satisfactions, 4 primary channels, and 3 maximum bandwidths.

These state definitions contain key parameters as for the Wi-Fi domain knowledge: \textit{i}) being satisfied or not is critical to user experience, and \textit{ii}) the spectrum configuration (or action) is most likely affecting the BSS performance. As for the former, being satisfied is actually the goal of the RL problem, so it will highly condition the action selection. Similarly, the current action seems a good indicator for suggesting which the next one should be.
Notice that higher multi-dimensional state definitions considering parameters like spectrum occupancy statistics would exponentially increase the state space, thus being cumbersome to explore and consequently reducing the learning speed dramatically.

To assess the convenience of states or contexts we use Q-learning and a contextual formulation of $\epsilon$-greedy. In Q-learning \eqref{eq:qlearning}, the learning rate $\alpha$ determines to what extent newly acquired information overrides old information: $\alpha = 0$ makes the agent learn nothing (exclusively exploiting prior knowledge) and $\alpha = 1$ makes the agent consider only the most recent information (ignoring prior knowledge to explore possibilities). We use a high value of $\alpha = 0.8$ because of the non-stationarity of the multi-agent setting. The discount factor $\gamma$ determines the importance of future rewards: $\gamma=0$ will make the agent myopic (or short-sighted) by only considering current rewards, while $\gamma \rightarrow 1$ will make it strive for a long-term high reward. We use a relatively small $\gamma = 0.2$ to foster exploitation (for rapid convergence) in front of exploration.

Another way in which agents use partial information gathered from the environment is in the form of contexts. In plain, contexts can be formulated as \textit{states for stateless approaches}. That is, a contextual MAB can be instantiated like a particular MAB running separately per context. The learning is separated from one context to the other so that no information is shared between contexts like it is the case for Q-learning and other state-full approaches. In this case, we define the contexts as the states and propose having a separate MAB instance for each of the contexts.

\subsubsection{Evaluation}

We assume each BSS having the same traffic load, $\ell = 50$ Mbps. So, this experiment's long-term dynamism is due exclusively to the multi-agent setting. That is, the inner traffic load conditions do not vary, but the spectrum management configurations (or actions) do. For the sake of assessing the learning speed, we consider a relatively short time-horizon of 200 iterations, each of 5 seconds. We run 100 random simulations for each RL algorithm.

\begin{figure}[h]
\centering
\begin{subfigure}{.21\textwidth}
  \centering
  \includegraphics[width=1\linewidth]{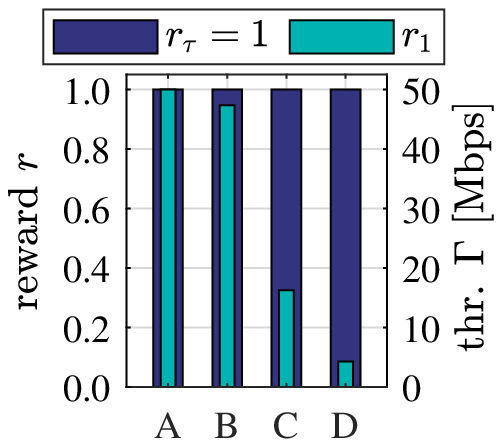}
  \caption{Random seed \#1 ($\tau = 31$).}
  \label{fig:seed1992}
\end{subfigure}%
~
\begin{subfigure}{.21\textwidth}
  \centering
  \includegraphics[width=1\linewidth]{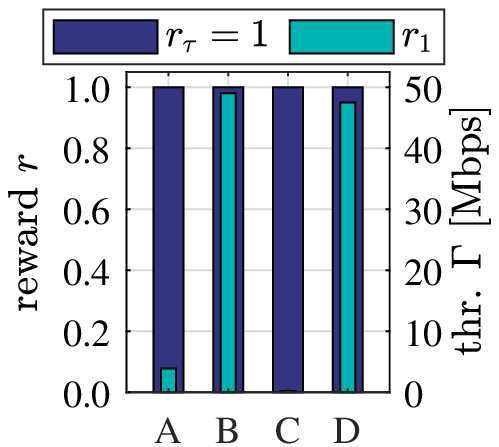}
  \caption{Random seed \#2 ($\tau = 28$).}
  \label{fig:seed2048}
\end{subfigure}
\caption{Initial reward $r_1$ (at $t=1$) and optimal reward $r_\tau = 1$, achieved in $t=\tau$ for two example random seeds. All BSSs use an $\epsilon$-greedy MAB formulation.}
\label{fig:seeds}
\end{figure}

Before comparing the performance of the different RL approaches, let us exhibit the need for configuration tuning of the primary and maximum bandwidth attributes in channel bonding WLANs. To that aim, we plot in Fig.~\ref{fig:seeds} the reward $r_1$ at the first iteration ($t=1$) and the optimal reward $r_\tau = 1$, where $\tau$ is the iteration when all the BSS reach optimal reward. All BSSs use an $\epsilon$-greedy MAB formulation. We plot such rewards for two example random seeds, i.e., for two random instantiations with random initial configurations. Since our reward definition is directly proportional to the throughput $\Gamma$, we also plot in the right y-axis the throughput in Mbps. Notice that in both cases, all the BSSs can reach the optimal reward at some point. However, the first instantiation (left subfigure) takes $\tau = 31$ iterations, whereas the second instantiation takes $\tau = 28$ iterations to reach the optimal, respectively.

\begin{figure}[t]
    \centering
    \includegraphics[width=0.85\linewidth]{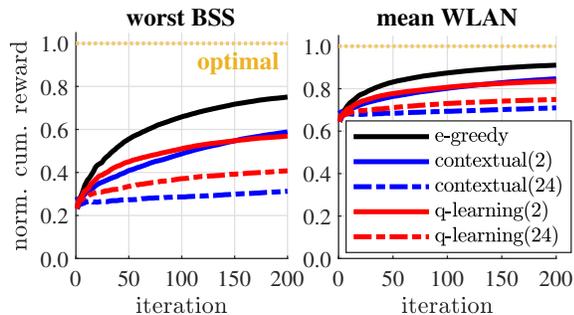}
    \caption{Run chart of the normalized cumulative gain for the worst BSS and WLAN's mean, averaged over 100 simulations. The number within the brackets in the contextual and Q-learning terms refers to the context/state space size.}
    \label{fig:evaluation}
\end{figure}

We now compare the different RL approaches presented. Figure~\ref{fig:evaluation} shows the normalized cumulative reward $G/t$ averaged through the 100 simulations for the worst-performing BSS and the WLAN's mean. We also show the optimal value, which turns out to be 1 for the worst and mean value since there exist a few global configurations where all the BSSs are satisfied. We plot results for the different RL approaches: raw $\epsilon$-greedy, contextual $\epsilon$-greedy for the binary and combined contextual spaces, and Q-learning also for the binary and combined state spaces. Notice that even though all the BSSs may reach the optimal at some point in a given simulation (as in Figure~\ref{fig:seeds}), they normally lose such optimal few iterations after since the considered algorithms do not take into account whether the individual (nor the global) optimal is reached. Therefore, any action change after reaching global optimality may affect the reward of all the BSSs.

We observe that while $\epsilon$-greedy converges (learns) to higher reward values close to the optimal, the other context/state-aided algorithms learn much more slowly, considerably below $\epsilon$-greedy's performance. In particular, the $\epsilon$-greedy MAB is able to get a near-optimal mean performance (just 7\% away from complete satisfaction). This relates to the fact that learning based on states/contexts is fruitless in dynamic and chaotic multi-agent deployments like this. So, it seems preferable to act quickly with a lower level of world awareness.
Besides, we note that the MAB is able to establish win-win relations between the BSSs, as shown by the worst performing BSS subplot. That is, on average, BSSs tend to benefit from others' benefits, resulting in fairer configuration. In particular, it outperforms the worst-case performance more than 30\% over the other algorithms.

Furthermore, we observe that both Q-learning and the contextual MAB formulation perform better when relying on the binary state/context space definition. That is, by shrinking the state space from 24 to 2 the learning speed increases. So, rather than aiding the learning, extensive state spaces slow it down, needing much more exploration. Besides, by eliminating the states and relying on the MAB, performance is significantly improved. This use-case showcases the MAB's ability to learn in multi-agent Wi-Fi deployments by establishing win-win relationships between BSSs.

%% file: sections/future.tex
\section{Concluding remarks}

We postulate MABs as an efficient ready-to-use formulation for decentralized spectrum allocation in multi-agent DCB WLANs. We argue why complex RL approaches relying on states do not suit the problem in terms of fast adaptability.
In this regard, we envision the use of states for larger time horizons, when the focus is not put on rapid-adaptation but on reaching higher potential performance in the long-run.